\documentclass[aps,prl,amsfonts,amssymb,
twocolumn,groupedaddress]{revtex4}

\usepackage{graphicx}

\begin{document}

\title{Dissociation of Vortex Stacks into Fractional-Flux 
   Vortices}
\author{A.\ De Col$^{\, a}$, V.B.\ Geshkenbein$^{\, a,b}$, 
and G.\ Blatter$^{\, a}$}
\affiliation{$^{a}$Theoretische Physik, ETH-H\"onggerberg, 
   CH-8093 Z\"urich, Switzerland}
\affiliation{$^{b}$L.D.\ Landau Institute for Theoretical Physics RAS,
   117940 Moscow, Russia}

\date{\today}

\begin{abstract}

We discuss the zero field superconducting phase transition in 
a finite system of magnetically coupled superconducting layers.  
Transverse screening is modified by the presence of other
layers resulting in topological excitations with fractional
flux. Vortex stacks trapping a full flux and present at any
finite temperature undergo a dissociation transition which 
corresponds to the depairing of fractional-flux vortices 
in individual layers. We propose an experiment with a 
bi-layer system allowing to identify the dissociation 
of bound vortex molecules.

\end{abstract}

\maketitle

The zero field superconducting to normal transition in thin films
and layered superconductors is triggered by the proliferation 
of topological defects; the unbinding of Pearl vortices 
\cite{pearl64} in thin films and of pancake vortices 
\cite{clem91} in layered superconductors generates a 
Berezinskii-Kosterlitz-Thouless transition (BKT) \cite{bkt73} 
which has been studied in detail \cite{kadinfiory83,artemenko89}.
New features emerge when going to a system with a finite 
number $N$ of magnetically coupled layers. Besides Pearl 
type vortex stacks penetrating through the full array 
of layers, cf.\ Fig.\ 1(a), fractional-flux vortices 
appear which reside within the individual layers
\cite{pudikov93,mints00}, the analogue of the pancake 
vortices in a layered material; the reduced trapped flux 
associated with these vortices is due to the presence 
of other layers modifying transverse screening in
the multi-layer system. The Pearl vortex can be viewed 
as a linear arrangement (stack) of fractional-flux vortices; 
the intra-layer unbinding transition of these fractional-flux 
vortices then corresponds to the dissociation of full-flux
vortex stacks present at any temperature, cf.\ Fig.\ 1(b). 
In this letter, we discuss the prospects of observing 
this dissociation transition in an experiment; in particular, 
we study a bi-layer system in a counterflow geometry 
which allows to observe the dissociation of vortex 
molecules into half-flux vortices, cf.\ Fig.\ 1(c). 
\begin{figure}[h]
\centering \includegraphics [width=8.6 cm] {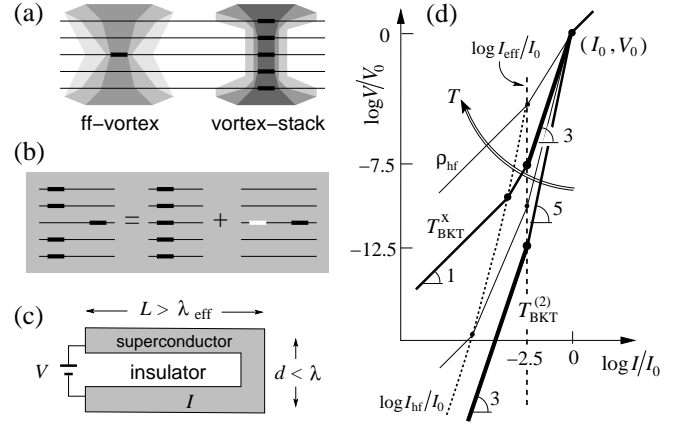}
\caption{(a) Fractional flux vortex and vortex stack in
a $N=5$-layer system. (b) The excursion of one ff vortex 
from the vortex stack is equivalent to the combination of 
a complete stack and a vortex--anti-vortex pair; pair 
unbinding in a layer is equivalent to stack dissociation.
(c)  Geometry for the bi-layer setup shortcircuiting the
effect of vortex stacks. (d) Sketch of $I$-$V$ curves at
various temperatures (see arrow); the algebraic dependence 
$V \propto I^\alpha$ with $\alpha>3$ at low temperatures 
turns ohmic ($\alpha=1$) at high temperatures. The 
regime between $T_{\rm \scriptscriptstyle BKT}^{\rm 
\scriptscriptstyle (2)}$ and $T_{\rm\scriptscriptstyle
BKT}^\mathrm{x}$ contains the interesting features
associated with half-flux vortices. The current scale
$I_\mathrm{eff}\sim I_0 (\xi/\lambda_\mathrm{eff})$
(vertical dashed line) separates the physics of unscreened
vortices from that of half-flux vortices. At $T_{\rm
\scriptscriptstyle BKT}^{\rm\scriptscriptstyle (2)}$
the $I$-$V$ curve exhibits the characteristic exponents
3 and 5 at small and large currents tracing the crossover 
from unscreened to half-flux vortices. Above $T_{\rm
\scriptscriptstyle BKT}^{\rm \scriptscriptstyle (2)}$ 
an additional ohmic regime due to free half-flux 
vortices appears at low currents; the dotted line 
marks the crossover current $I_\mathrm{hf}$.  At 
temperatures $T>T_{\rm\scriptscriptstyle BKT}^\mathrm{x}$
the ohmic regime takes over and leaves only a small
non-linear region at high currents probing unscreened
vortices.}
\label{fig:iv_setup}
\end{figure}

The basic prerequisite for the appearance of a BKT
transition is the logarithmic interaction between defects, 
$V(R) = 2e^2 \ln R$, where we attribute an effective 
`charge' $e$ to the defects. In the absence of screening,
e.g., in a superfluid $^4$He film, the logarithmic 
interaction between vortices extends to infinity and 
the system undergoes a finite temperature BKT transition 
\cite{bkt73}. In a superconducting film, transverse 
screening restricts the log-interaction to the
screening length $\Lambda$; the Pearl vortices assume 
a finite self-energy $V(\Lambda)/2$ and hence can 
be thermally excited at any finite temperature --- 
the superconducting to normal transition then is 
shifted to $T=0$, although a sharp crossover survives 
at a finite temperature $T^\mathrm{x}_\mathrm{\rm 
\scriptscriptstyle BKT} \approx e^2/2$ \cite{kadinfiory83}.
Below, we concentrate on a system with $N$ magnetically
coupled superconducting layers, i.e., vortices interact
through the transverse magnetic potential ${\bf A}$,
while the Josephson coupling due to Cooper pair tunneling 
between the layers is assumed to be negligible, as it is the 
case in a material with insulating layers separating the 
superconducting ones. The presence of additional layers 
leads to drastic modifications in the potential $V(R)$ 
between individual vortices in the same layer: {\it i)} 
the log-interaction extends to infinity, 
{\it ii)} the magnetic flux $\Phi_\mathrm{t}$ 
trapped by individual vortices is reduced to a fraction 
$1/N$ of the flux unit $\Phi_0 = hc/2e$, $\Phi_\mathrm{t}
(N) = \Phi_0/N$; the effective coupling $e^2$ 
in the interaction potential between defects is
reduced correspondingly, $e^2 \rightarrow e^2 [1-1/N]$. 
Hence, the additional layers act on the vortices in the 
same way as a dielectric matrix acts on charged particles. 
The unbinding of the fractional-flux (ff) vortices triggers a 
finite temperature BKT transition at $T^{\scriptscriptstyle 
(N)}_\mathrm{\rm \scriptscriptstyle BKT} \approx e^2 
[1-1/N]/2$. At the same time, vortex stacks are present 
at any temperature; the unbinding of (ff) vortices at 
$T^{\scriptscriptstyle (N)}_\mathrm{\rm \scriptscriptstyle 
BKT}$ then describes the dissociation of the vortex stacks 
rather than the superconductor-normal transition which, 
strictly speaking, appears already at $T=0$. 
Finally, in a bulk layered superconductor, the magnetic 
field escapes in the transverse direction and the 
flux trapped by pancake vortices saturates at
$\Phi_\mathrm{t} = d\Phi_0/ 2\lambda$, where $d$ and
$\lambda$ denote the layer separation and the bulk planar
penetration depth. The pancake-vortex unbinding at
$T_\mathrm{\rm \scriptscriptstyle BKT}^{\scriptscriptstyle 
(\infty)} \approx e^2/2$ describes a generic phase transition 
as no competing stacks show up in the bulk. 

Below, we address the prospect of observing fractional-flux
vortices in an experiment. The presence of vortices
and their unpairing reveals itself in the current-voltage 
characteristic. Fractional-flux vortices appear most 
prominently in a bi-layer system, see Fig.\ 1(c). 
We propose a counterflow experiment where the 
contribution of stacks (vortex molecules) is eliminated. 
The current-induced unpairing of vortices then produces an 
algebraic characteristic $V \propto I^\alpha$; the change
in the exponent from $\alpha = 3$ to 5 with increasing 
current, cf.\ Fig.\ 1(d), is a manifestation of the stack 
dissociation transition at $T^{\scriptscriptstyle
(2)}_\mathrm{\rm \scriptscriptstyle BKT}$. In the following, 
we briefly derive the structure of topological excitations 
in layered systems, discuss their thermodynamics, and 
analyze the features in the $I$-$V$ characteristic 
related to fractional-flux vortices.

Consider a superconductor of thickness $d$ and with a 
London penetration depth $\lambda$. Central to our 
discussion is the interaction potential $V(R)$ between 
vortices: The current associated with a vortex is driven 
by the $2\pi$-phase twist $\nabla\varphi= -\hat{\bf n}_z
\times {\bf R}/R^2$, ${\bf j}(R)=-(c/4\pi \lambda^2) 
[\nabla \varphi \, \Phi_0/2\pi+{\bf A}(R)]$; transverse 
screening through the vector potential ${\bf A}$ reduces 
the action of $\nabla \varphi$ until complete compensation 
is reached once a full flux quantum $\Phi_0$ is trapped. 
A second vortex placed a distance $R$ away is subject to 
the Lorentz force $F(R)= -j_{\phi}(R) d \Phi_0/c = -(2
\varepsilon_0 d/R)[1-\Phi(R)/ \Phi_0]$, where $\Phi(R)
=2\pi R A_{\phi}(R)$ is the flux accumulated within the 
distance $R$ and $\varepsilon_0 = (\Phi_0/4\pi \lambda)^2$
is the line energy; integrating this force provides us 
with the desired potential between the vortices. The 
incomplete asymptotic screening with a reduced trapped 
flux $\Phi_\mathrm{t} \equiv \Phi(R\to\infty) < \Phi_0$ 
then gives rise to a logarithmic interaction $V(R)\sim 
2\varepsilon_0 d [1-\Phi_\mathrm{t} /\Phi_0] \ln (R)$ 
and hence a BKT phase transition (note that $e^2 
\leftrightarrow \varepsilon_0 d$). 

In order to find the flux $\Phi_\mathrm{t}$ 
we have to solve the Maxwell equations for the potential 
${\bf A}$. We consider a stack of $N$ superconducting
layers of thickness $d_\mathrm{s}$ and separated by a 
distance $d$. The penetration depth $\lambda_\mathrm{s}$
of the layer material defines the bulk planar 
penetration depth $\lambda^2 = \lambda_\mathrm{s}^2
d/d_\mathrm{s}$. We place the vortex at the origin of
the layer positioned at $z=0$ and describe the protecting
layers of thickness $d^{\rm\scriptscriptstyle <} =(N-n)d$ and
$d^{\rm\scriptscriptstyle >}=(n-1)d$ above and below the film
in a continuum approximation. The system 
\begin{eqnarray}
   &&\nabla^2 {\bf A}\!-\!\frac{1}{\lambda^2}{\bf A}
    = \frac{d}{\lambda^2}\Bigl({\bf A}
    \!+\!\frac{\Phi_0}{2\pi}{\bf\nabla}\varphi\Bigr)\delta(z),
    \quad\!\!-d^{\rm\scriptscriptstyle <}\!<z\!<\!
              d^{\rm\scriptscriptstyle >}\!\!, 
   \nonumber \\ 
   &&\nabla^2 {\bf A} = 0,
   \quad\quad\quad\quad z<-d^{\rm\scriptscriptstyle <}
   \mbox{ and  }           z> d^{\rm\scriptscriptstyle >}
   \label{a3}
\end{eqnarray}
then assumes the solution 
\begin{equation}\label{3d}
   A_{\phi}(R,z)
   =\frac{\Phi_0 d}{\lambda^2}\int_0^\infty\frac{dK}{2\pi}
   \frac{J_1(KR)}{C(K)} f(K,z),
\end{equation}
with $J_1$ the Bessel function and the function $f(K,z) =
[1-\alpha_{d'}(K)]e^{-K_+ |z|}
+\alpha_{d'}(K)e^{K_+|z|}$ describing the $z$-dependence {\it
within the superconductor}. Here, $\alpha_{d'}(K)=
(K_+-K)/[(K_++K)e^{2K_+d'}+(K_+-K)]$, with $K_+=\sqrt{K^2
+\lambda^{-2}}$ and $d'=d^{\scriptscriptstyle <}$ ($d' =
d^{\scriptscriptstyle >}$) in the region
$-d^{\scriptscriptstyle <}<z<0$ ($0<z<d^{\scriptscriptstyle
>}$).  The denominator $C(K)$ assumes the form
$C(K)=[1-2\alpha_{d^<}(K)] K_+ + [1-2\alpha_{d^>}(K)] K_++
d/\lambda^2$. {\it Outside the superconductor}, the field is
obtained by replacing $z>0$ ($z<0$) by $d^{\scriptscriptstyle
>}$ ($-d^{\scriptscriptstyle <}$) in $f$ and an additional
correction factor $\exp[K (d^{\scriptscriptstyle >}-z)]$
($\exp[K(d^{\scriptscriptstyle <} +z)]$). The Pearl- and
pancake vortices are recovered in the limits
$d^{\scriptscriptstyle <},\, d^{\scriptscriptstyle >} =0$ and
$d^{\scriptscriptstyle <},\, d^{\scriptscriptstyle >}=\infty$.

The magnetic flux $\Phi_\mathrm{t}$ trapped  by a vortex
is extracted from the vector potential at $z=0$; for a thin 
$N$-layer system with $Nd \ll \lambda$ we find the asymptotic 
form
\begin{equation}\label{a_asymp2}
   A_{\phi}(R\gg\lambda_{\mathrm{eff}},z=0)
   \sim (\Phi_\mathrm{t}/ 2\pi R) (1-\lambda_\mathrm{eff}/R)
\end{equation}
with parameters $\lambda_{\mathrm{eff}} \approx 2 \lambda^2/d
N$ and $\Phi_\mathrm{t}(N)\approx \Phi_0/N$ independent of the
layer position $n$. For large $N$ the trapped flux saturates
at the value $\Phi_{\mathrm{t}} = d\Phi_0/2\lambda$ and
$\lambda_{\mathrm{eff}} = 0$ \cite{corr}, while
$\Phi_{\mathrm{t}}=\Phi_0$ and $\lambda_{\mathrm{eff}}
=\Lambda = 2\lambda_\mathrm{s}^2/ d_\mathrm{s}$ for the thin
film. The decrease $\Phi_\mathrm{t} = \Phi_0/N$ in trapped
flux with increasing number of layers is easily understood:  
with $Nd \ll\lambda$, the same magnetic field (and hence 
the same flux) is penetrating the $N$ layers. On the other 
hand, a vortex stack (i.e., $N$ (ff) vortices) 
carries a full flux $\Phi_0$. The flux associated
with one individual (ff) vortex then is a fraction $1/N$ of
the value trapped by the vortex stack and thus
$\Phi_\mathrm{t} = \Phi_0/N$.

The incomplete screening of the vortex singularity produces 
a log-interaction between vortices, $V(R) = 2\varepsilon_0d 
\ln(R/\xi)$ at small distances $R\ll\lambda_{\mathrm{eff}}$ 
and $V(R) = 2\varepsilon_0d [\ln(\lambda_\mathrm{eff}/\xi)
+[1-\Phi_{\mathrm{t}}/\Phi_0] \ln(R/\lambda_\mathrm{eff})]$
involving a large self-energy but a reduced prefactor 
at large distances $R \gg \lambda_{\mathrm{eff}}$. This 
logarithmic interaction competes with the entropy of vortex-pair 
excitations and triggers a Berezinskii-Kosterlitz-Thouless 
transition at \cite{renorm,pudikov93}
\begin{equation}
   T^{\scriptscriptstyle (N)}_{\rm\scriptscriptstyle BKT}
   =\frac{\tilde{\varepsilon}_0 d}{2}
   \Bigl(1-\frac{\Phi_\mathrm{t}(N)}{\Phi_0}\Bigr)
   \label{bkt}.
\end{equation} 
In a thin film, $\Phi_\mathrm{t} = \Phi_0$ and the finite
range of the interaction between Pearl vortices pushes the
real transition to $T=0$, in agreement with (\ref{bkt});  
the presence of one additional `protecting' layer changes the
situation: transverse screening reduces the trapped flux to
half its value, $\Phi_\mathrm{t} = \Phi_0/2$, thus extending
the range of the logarithmic interaction to infinity and
pushing the transition temperature to a finite value
$T^{\scriptscriptstyle (2)}_{\rm\scriptscriptstyle BKT} =
\tilde{\varepsilon}_0 d/4$. Adding more ($N-1$) layers, the
trapped flux $\Phi_\mathrm{t} = \Phi_0/N$ decreases further
until assuming the asymptotic value $\Phi_\mathrm{t}\!=
\!d\Phi_0/2\lambda$ in a bulk superconductor where 
$T_{\rm\scriptscriptstyle BKT}^{\scriptscriptstyle (\infty)} 
\approx \tilde{\varepsilon}_0 d/2$ is largest.

The appearance of a finite temperature phase transition due to
the protecting action of additional superconducting layers has
its counterpart in multigap superconductors \cite{babaev02}, 
cf.\ also Ref.\ \onlinecite{sachdev92sudbo04}. In both cases, 
the creation of a topological defect in one superconducting 
component or layer induces screening currents in the other 
components/layers via coupling to the common gauge field 
${\bf A}$; the resulting incomplete screening extends the 
interaction between defects to infinity, although with a 
reduced `charge'. A finite Josephson coupling 
between the layers or between the components of a multigap 
superconductor spoils this phenomenon through the appearance 
of a linear confining potential. While this coupling can 
be (made) arbitrarily small in a layered system, the internal 
Josephson effect in a multi-component superconductor cannot 
be tuned and is not necessarily small \cite{gorokhov04}.

The setup where fractional-flux vortices make their most 
prominent appearance is the bi-layer system. Its thermodynamic 
properties are obtained from an extension of the renormalization 
group analysis in a film \cite{k74,horovitz93} and 
involves the flow of the superfluid density $K(R)$ at scale 
$R$ with $K(\xi) \equiv K_0 =\varepsilon_0 d/\pi T$ and 
the vortex fugacity $y(R)$, with $(y/R)^2$ the density of 
vortex--anti-vortex pairs of size $R$. Here, the 
renormalization involves a two-stage process: {\it i)} 
unscreened vortex pairs are integrated out on scales 
$R<\lambda_\mathrm{eff}$ and provide renormalized values 
$(\bar{K}, \bar{y})$ at $R= \lambda_\mathrm{eff}$; {\it ii)} 
the flow is restarted with a reduced coupling $\varepsilon_0 
d/2$ and half-flux (hf) vortex pairs are integrated out 
on scales $R> \lambda_\mathrm{eff}$.  We obtain the 
following results (see \cite{decol04} for details): 
At temperatures $T < T_{\rm\scriptscriptstyle BKT}^{\rm 
\scriptscriptstyle (2)}=\tilde{\varepsilon}_0 d/4$ the 
initial fugacity $y_0=\exp(-E_c/T)$ due to the core energy
$E_c \sim \varepsilon_0 d$ flows to zero and the
superfluid density $K\approx\bar{K}> 4/\pi$ remains finite.
Above $T_{\rm \scriptscriptstyle BKT}^{\rm\scriptscriptstyle
(2)}$ (hf) vortices start unbinding: a narrow critical
regime (with a correlation length $\xi_\mathrm{hf}\approx
\lambda_\mathrm{eff} \exp(\pi/2\sqrt{b \Delta t})$, $\Delta t 
\equiv (T-T_{\rm\scriptscriptstyle BKT}^{\rm\scriptscriptstyle
(2)})/T_{\rm \scriptscriptstyle BKT}^{\rm\scriptscriptstyle
(2)}$ and $b$ a dimensionless parameter) is followed by a 
mean-field behavior where the fugacity diverges to 
infinity and the renormalized superfluid density vanishes 
beyond the correlation length $\xi_\mathrm{hf}$, 
defining a density of free (hf) vortices
\begin{equation}\label{hf}
    n_{\mathrm{hf}}\approx\frac{1}{\xi^2}
    \Bigl(\frac{\xi y_0^{2/\pi\bar{K}}}
               {\lambda_\mathrm{eff}}\Bigr)
    ^{\textstyle{\frac{2\pi\bar{K}}{4-\pi\bar{K}}}}
    =\frac{\lambda_{\mathrm{eff}}^2}{\xi^4}
    \Bigl(\frac{\xi \sqrt{y_0}}{\lambda_\mathrm{eff}}\Bigr)
    ^{\textstyle{\frac{a}{\Delta t}}},
 \end{equation}
with $a$ of order unity. On approaching $T_{\rm
\scriptscriptstyle BKT}^\mathrm{x} =\tilde{\varepsilon}_0 
d/2$ the correlation length becomes comparable to 
$\lambda_\mathrm{eff}$; beyond $T_{\rm 
\scriptscriptstyle BKT}^\mathrm{x}$ unscreened
vortices start unbinding at small scales below 
$\lambda_\mathrm{eff}$. Note that vortex stacks 
do preempt the superconducting transition of the 
bi-layer system but preserve superconductivity 
within the individual layers; indeed, the force of 
a (hf) vortex acting on a vortex stack vanishes 
rapidly beyond the effective screening length
$\lambda_\mathrm{eff}$ due to the complete 
screening of vortex stacks. 

The presence of (hf) vortices can be traced in an
experiment measuring the $I$-$V$ characteristic (we denote the
sheet-current density by $I = j d$). In the counterflow
geometry of Fig.\ 1(c), the applied $dc$ current acts
oppositely on the two (hf) vortices constituting a stack
and the linear response due to drag motion is quenched.
The current-induced dissociation of pairs and stacks of 
(hf) vortices produces a non-linear $I$-$V$ characteristic;
the change in slope from 3 at low currents to 5 at high
currents signals the thermodynamic dissociation of stacks
at $T_{\rm\scriptscriptstyle {BKT}}^{\rm\scriptscriptstyle 
(2)}$. The current-induced unbinding of (hf) vortex pairs 
involves a thermal activation over the barrier $U(I) = 
\max_R [V(R)-I\Phi_0 R/c] \approx 2\tilde{\varepsilon}_0 d 
\ln(R_I/\xi)$ at small distances $R_I\ll \lambda_\mathrm{eff}$, 
while $U \approx 2\tilde{\varepsilon}_0 d \ln(\lambda_\mathrm{eff}/\xi)
+\tilde{\varepsilon}_0 d\ln(R_I/\lambda_\mathrm{eff})$ for
$R_I \gg \lambda_\mathrm{eff}$. Here, $R_I \approx \xi I_0/I$
denotes the unbinding scale and $I_0=2\tilde{\varepsilon}_0 d
c/\Phi_0\xi$ is close to the depairing current. Applied currents
smaller than $I_\mathrm{eff}=I_0 \xi/\lambda_{\mathrm{eff}}$
probe lengths larger than $\lambda_\mathrm{eff}$ and the
effects of half-flux vortices become accessible.

The equilibrium density $n_\mathrm{v}$ of free vortices
derives from the steady state solution of the rate equation
\cite{halperinnelson79} $\partial_t n_\mathrm{v} =\Gamma-
\xi^2 n_\mathrm{v}^2/\tau_{\rm rec}$, with
$\Gamma\propto\exp(-U/T)$ the production rate of free vortices
and $\xi^2/\tau_{{\rm rec}}$ the recombination parameter.
Vortex drag then produces a finite non-linear resistivity,
$\rho \approx \xi^2 \rho_\mathrm{n} n_\mathrm{v}$ with
$\rho_\mathrm{n}$ the normal state resistivity, and a
corresponding algebraic $I$-$V$ characteristic $V/V_0 \sim
(I/I_0)^{\alpha(I)}$ with
\begin{equation}
   \alpha(T,I)=1+\pi\bar{K}(T)[1-\Phi_\mathrm{t}(R_I)/\Phi_0].
   \label{alpha}
\end{equation}
The exponent $\alpha$ depends explicitly on the flux
associated with the vortices: at short scales, unscreened
vortices are probed and $\alpha=1+ \pi\bar{K}$.  On
the other hand, large distances probe half-flux vortices and
the exponent is reduced to $\alpha=1+\pi \bar{K}/2$. The
crossover between these two regimes appears at the current
$I_\mathrm{eff} \ll I_0$. This reduction in $\alpha$ at
$I_\mathrm{eff}$ is the most prominent feature in the $I$-$V$
characteristic signalling the presence of half-flux vortices 
in the system; at $T_{\rm\scriptscriptstyle BKT}^{\rm
\scriptscriptstyle (2)}$, the change in slope is from 5 at large 
currents to 3 at low currents, cf.\ Fig.\ 1(d). The associated 
voltage signals are weak as the density of free vortices is 
already small, $n_\mathrm{v} \sim 1/\lambda_\mathrm{eff}^2$ 
at $T_{\rm\scriptscriptstyle BKT}^\mathrm{x}$ and $n_\mathrm{v} 
\sim \xi^2/ \lambda_\mathrm{eff}^4$ at the true transition 
point $T_{\rm\scriptscriptstyle BKT}^{\rm\scriptscriptstyle (2)}$.  

The temperature $T_{\rm\scriptscriptstyle {BKT}}^{\rm
\scriptscriptstyle {(2)}}$ defines a resistive transition \
due to the proliferation of free half-flux vortices. Above
$T_{\rm\scriptscriptstyle {BKT}}^{\rm \scriptscriptstyle (2)}$
the ohmic resistance appearing at low currents $I <
I_\mathrm{hf} = I_0\xi/\xi_\mathrm{hf}$ (probing distances
larger than the correlation length $\xi_\mathrm{hf}$) is
determined by the density of free (hf) vortices, $\rho \approx
\rho_\mathrm{n} \xi^2 n_\mathrm{hf}$. In the mean-field regime
above $T_{\rm\scriptscriptstyle BKT}^{\rm \scriptscriptstyle
(2)}$ we can make use of (\ref{hf}) and find the location of
the crossover at $I_\mathrm{hf}\approx I_0
(\lambda_{\mathrm{eff}}/\xi) (\xi\sqrt{y_0}/
\lambda_\mathrm{eff})^{a/2\Delta t}$ and the temperature
dependent resistivity $\rho_\mathrm{hf}(T)  \approx
\rho_\mathrm{n} (\lambda_{\mathrm{eff}}/\xi)^2 (\xi
\sqrt{y_0}/\lambda_{\mathrm{eff}})^{a/\Delta t}$ due to free
half-flux vortices. The measurement of $I_\mathrm{hf}$ or
$\rho_\mathrm{hf}$ in this regime provides direct access to the
correlation length $\xi_\mathrm{hf}$ and its mean-field like
temperature dependence.

In order to analyze the BKT transition at $T_{\rm
\scriptscriptstyle {BKT}}^{\rm\scriptscriptstyle (2)}$
various experimental constraints have to be accounted for:
{\it i)} The system must be larger than the screening length
$\lambda_\mathrm{eff}$ beyond which (hf) vortices appear.
The relation (\ref{bkt}) implies that $\lambda_\mathrm{eff}
(T_{\rm\scriptscriptstyle BKT}^{\rm \scriptscriptstyle (2)})
\approx 0.5~{\rm cm}/(T_{\rm\scriptscriptstyle BKT}^{\rm
\scriptscriptstyle (2)} \mathrm{~in~K})$, hence
$\lambda_{\mathrm{eff}} \approx 1$ mm in a typical low $T_{\rm
c}$ material.  {\it ii)} Both the interlayer distance $d$ and
the layer thickness $d_s$ have to be small compared to the
bulk penetration depth $\lambda$. {\it iii)} The Josephson
coupling has to be small enough to push the confinement length
$\Lambda_\mathrm{c} = \sqrt{(j_0/j_J) \xi d}$ beyond 
$\lambda_\mathrm{eff}$; separating the superconducting
layers with an insulator \cite{thick} the Josephson current 
$j_J$ can be made arbitrarily small. {\it iv)} The mean-field 
temperature dependence $\propto (1-T^2/T_\mathrm{c}^2)^{-1/2}$ 
of the parameters $\lambda$ and $\xi$ tends to push the 
temperatures $T_{\rm\scriptscriptstyle BKT}^{\rm \scriptscriptstyle 
(2)}$ and $T_{\rm\scriptscriptstyle BKT}^\mathrm{x}$ towards 
the mean-field critical temperature $T_c$, $(T_{\rm
\scriptscriptstyle BKT}^\mathrm{x} - T_{\rm\scriptscriptstyle 
BKT}^{\rm\scriptscriptstyle (2)})/ T_c \approx (T_\mathrm{c} 
- T_{\rm\scriptscriptstyle BKT}^\mathrm{x})/T_\mathrm{c} 
\approx 4 Gi^{\rm\scriptscriptstyle {(2D)}}$ with $Gi^{\rm
\scriptscriptstyle {(2D)}}= T_c/2\varepsilon_{0}(T=0)d\ll 1$ 
the two-dimensional Ginzburg number \cite{review94}. A large 
Ginzburg number helps in distinguishing between 
the temperatures where (hf) and unscreened 
vortices unbind. {\it v)} The features in the $I$-$V$ 
characteristic identifying the presence of (hf)
vortices involve vortex densities which are suppressed by 
the small parameter $\xi/\lambda_\mathrm{eff}$. Correcting 
parameters for the intrinsic dirtiness of thin films (see 
\cite{deGennes66}, we assume a mean free path $l$ limited 
by the layer thickness $d_s$) we obtain the estimates 
\begin{eqnarray}
  [\xi/\lambda_\mathrm{eff}]_{T_{\rm
  \scriptscriptstyle BKT}^{\rm \scriptscriptstyle (2)}} 
  \approx 
  1.5\cdot 10^{-4}\  T_\mathrm{c}^{1/2} d_s^{3/2}/
  \lambda_\mathrm{c0}, 
  \label{xileff}\\
  Gi^{\rm\scriptscriptstyle {(2D)}} 
  \approx 3.2\cdot 10^{-9}\  T_\mathrm{c}
  \, \lambda_\mathrm{c0}^2 \xi_\mathrm{c0}/d_s^2,
  \label{gi}
  \end{eqnarray}
where all lengths are measured in \AA\ and temperatures in
Kelvin. The results (\ref{xileff}) and (\ref{gi}) tell us that
given the (clean-) material parameters $\lambda_\mathrm{c0}$ 
and $\xi_\mathrm{c0}$ it is not possible to maximize 
both $Gi^{\rm\scriptscriptstyle (2D)}$ and $\xi/
\lambda_\mathrm{eff}$ simultaneously by varying the
thickness $d_s$. A reasonable compromise can be achieved if 
we choose a material with $\lambda_\mathrm{c0} \sim
\xi_\mathrm{c0} \sim 1000$ \AA, $T_\mathrm{c}\sim 10$ K, 
and a thickness $d_s\sim 500$ \AA; this yields $\xi/
\lambda_\mathrm{eff}\approx 10^{-2.5}$ and $Gi^{\rm
\scriptscriptstyle (2D)} \sim 10^{-4}$. The small
value of $\xi/ \lambda_\mathrm{eff}$ implies a small
vortex density and requires a high voltage resolution,
while the smallness of $Gi^{\rm \scriptscriptstyle 
(2D)}$ requires a temperature resolution in the mK range.
The characteristic halving in $\alpha-1$ signalling the
presence of (hf) vortices below $I_\mathrm{eff}$
involves voltages with $\log(V/V_0)$ between
$-7.5$ and $-12.5$. With $V_0 \sim \rho_\mathrm{n} 
j_0 L \approx 10~$ mV (we assume $\rho_\mathrm{n} 
\sim 100~\mu\Omega$cm, $j_0(T_{\rm \scriptscriptstyle 
BKT}^{\rm\scriptscriptstyle (2)}) \sim 10^2~\mathrm{A} 
/\mathrm{cm}^2$, and $L\sim 1~\mathrm{cm}$)
we find that an experimental voltage resolution in the
sub-pico-Volt regime \cite{gammel91} allows to trace
this halving in $\alpha-1$ over a substantial 
temperature range below $T_{\rm \scriptscriptstyle 
BKT}^\mathrm{x}$, although the observation of this 
crossover at $T_{\rm \scriptscriptstyle BKT}^{\rm
\scriptscriptstyle (2)}$ 
itself pushes the limits of present days experimental 
capabilities. Alternatively,  
one can trace the presence of 
(hf) vortices by measuring 
the characteristic  mean-field type resistivity 
$\rho_\mathrm{hf} (T)$ below $T_{\rm \scriptscriptstyle BKT}^\mathrm{x}$ and through direct observation with a scanning SQUID microscope \cite{tafuri04}.

An interesting analogy appears when comparing the present
system with the bi-layer quantum Hall setup at total filling 
$\nu=1$. The latter is expected to undergo a BKT transition 
into an interlayer phase coherent state, even in the
absence of any tunneling between the layers \cite{wenzee92}. 
The (hf) vortices discussed above (existing in one of two 
layers and with $\pm$ vorticity) correspond to topological
excitations (merons) with charge $\pm e/2$ and $\pm$
vorticity \cite{moon95}; bound neutral meron pairs have 
their analogue in intralayer vortex--anti-vortex pairs, 
while bound charged merons correspond to vortex stacks. 
The unbinding of meron-pairs in the BKT transition 
destroys the interlayer phase coherence and can be 
traced in the same type of counterflow experiment 
\cite{kellogg04} as discussed above.

We acknowledge financial support from the Swiss National
Science Foundation through the program MaNEP.


\begin{thebibliography}{99}
\bibitem{corr} In the bulk, corrections to the leading term 
   $A \sim \Phi_\mathrm{t}/2\pi R$ are exponentially small for 
   $R\gg\lambda$.
\bibitem{thick} With an exponential dependence on $d$, an 
   insulator of a few nm thickness is sufficient.
\end{thebibliography}
\end{document}